\documentclass[conference]{IEEEtran}
\IEEEoverridecommandlockouts

\usepackage{cite}
\usepackage{amsmath,amssymb,amsfonts}
\usepackage{algorithmic}
\usepackage{graphicx}
\usepackage{textcomp}
\usepackage{url}
\def\BibTeX{{\rm B\kern-.05em{\sc i\kern-.025em b}\kern-.08em
    T\kern-.1667em\lower.7ex\hbox{E}\kern-.125emX}}
\begin{document}

\onecolumn
\textcopyright 2018 IEEE. Personal use of this material is permitted. Permission from IEEE must be obtained
for all other uses, in any current or future media, including reprinting/republishing this material for advertising or promotional purposes, creating new collective works, for resale or redistribution to servers or lists, or reuse of any copyrighted component of this work in other works.
\newline
\newline
The final, published version of this paper is available under:
S. Marksteiner, "Reasoning on Adopting OPC UA for an IoT-Enhanced Smart Energy System from a Security Perspective,"
\textit{2018 IEEE 20th Conference on Business Informatics (CBI)}, Vienna, Austria, 2018, pp. 140-143. doi:
10.1109/CBI.2018.10060.
URL: \url{https://ieeexplore.ieee.org/document/8453946/}
\twocolumn

\title{Reasoning on Adopting OPC UA for an IoT-Enhanced Smart Energy System from a Security Perspective
\thanks{This work has received funding from the European Union's Horizon 2020 research and innovation programme under
grant agreement No 773715, Project RESOLVD (LCE-01-2016-2017).}
}

\author{\IEEEauthorblockN{Stefan Marksteiner}
\IEEEauthorblockA{
	\textit{DIGITAL - Institute for Information} \\
	\textit{and Communication Technologies}\\
	\textit{JOANNEUM RESEARCH GmbH}\\
	Graz, Austria\\ 
	Email: stefan.marksteiner@joanneum.at
	}
}

\IEEEspecialpapernotice{(Workshop Paper)}
                
\maketitle

\begin{abstract}
	Smart Services using \textit{Industrial Internet of Things (IIoT)} applications are on the rise, but still more often
than not, traditional industrial protocols are used to interconnect the entities of the resulting systems. These
protocols are mostly not intended for functioning in such a highly interconnected environment and, therefore, often lack even the
most fundamental security measures. To address this issue, this paper reasons on the security of a communications
protocol, intended for \textit{Machine to machine (M2M)} communications, namely the \textit{Open Platform Communications Unified Architecture
(OPC UA)} and exemplifies, on a smart energy system, its capability to serve as a secure communications
architecture by either itself or in conjunction with traditional protocols.

\end{abstract}


\section{Introduction and Motivation}
\label{sec:intro}

This paper aims on reasoning on the security of the \textit{Open Platform Communications Unified Architecture (OPC UA)}
for the use in a smart energy system (a controller and its managed devices). Concretely, it discusses whether using OPC
UA is an appropriate method to secure communications for an Internet of Things (IoT)-based architecture that allows
for advanced algorithms in low voltage distribution grids to improve efficiency and hosting capability. 
The original idea to provide this communications was via the in the industry widely proliferated protocol
\textit{Modbus/TCP} \cite{IECTR2014}.
This protocol, however, is known to have severe security deficiencies \cite{6016202}. It has therefore to be replaced or
security-enhanced by another protocol. Both tasks could be achieved by OPC UA, as it both provides a standalone
architecture for \textit{Machine-to-Machine Communication (M2M)} and there is work to run OPC UA with Modbus
\cite{6566647}.
The reason why OPC UA is deemed a suitable candidate for this type of communications is its standardization,
proliferation and also its semantic interoperability \cite{8104808}. This advanced flexibility and control will then
facilitate the advanced use of renewable energy, as well as new service-based business models in the energy domain. Assuring the security of
these communications, however,  is a fundamental prerequisite for this intended technology to work without generating
the risk of large-scale cyberattacks and the protocol's capabilities in that matter are therefore . 
 Another goal of this paper is to extend the
author's previous work on the security of IoT protocols \cite{Marksteiner:2017:8260940} by adding a higher-level, industrial-use specification that allows for industry 4.0 and
energy domain use cases.

\section{Related Work}

There is a security analysis of OPC UA commissioned by the  \textit{German Federal Office for Information Security
(BSI)} \cite{Damm2017}. This work, however, was conducted over the course of the year 2015 and does therefore not take
into account newer developments (see Section \ref{sec:SA:OPC}). In any case, as this work is quite comprehensive, it
servers as a starting point for this paper, apart from the official OPC UA specifications in particular
the security model \cite{OPC2015} and the profiles \cite{OPC2017}. 

\section{Protocol Description}
The \textit{Open Platform Communications Unified Architecture (OPC UA)} is a system architecture that is designed to
exchange command and control information between industrial sensors, actuators, control systems, \textit{Manufacturing Execution
Systems (MES)} and \textit{Enterprise Resource Planning (ERP) Systems} \cite{OPC2017a}. It therefore operates on all of
the four upper layers of the \textit{IEC 62264 Enterprise-control system integration} norms \cite{IECTR2013}.
The architecture specifies the following \cite{OPC2017a}:
\begin{itemize}
	\item The information model to represent structure, behaviour and semantics;
	\item The message model to interact between applications;
	\item The communication model to transfer the data between end-points;
	\item The conformance model to guarantee interoperability between systems.
\end{itemize}
In order to provide a platform-independent infrastructure logic to flexibilize host services for the \textit{Industrial
Internet of Things (IIoT)}, enabling \textit{Machine-to-Machine Communication (M2M)}, it provides both a
\textit{Client-Server} and a \textit{Publisher-Subscriber (PubSub)} model. To model the actual data it defines three
encodings:
\begin{itemize}
  \item The \textit{Extensible Markup Language (XML)} \cite{bray1997extensible};
  \item The \textit{JavaScript Object Notation (JSON)} \cite{RFC8259};
  \item A native, binary encoding (\textit{UA Binary}).
\end{itemize} 
It further defines some protocols to transfer the modeled data:
\begin{itemize}
  \item OPC UA via the \textit{Transmission Control Protocol (TCP)};
  \item Via the \textit{Hypertext Transfer Protocol Secure (HTTPS)};
  \item Via \textit{WebSockets}. 
\end{itemize}

The architecture is an advancement of the \textit{Open Platform Communications (OPC)}, formerly called
\textit{OLE for Process Control} model \cite{OPC2017a}. The latter  historically derives from the Microsoft
\textit{Object Linking and Embedding (OLE)/Component Object Model (COM)} \cite{1195286}.
The protocol has been internationally standardized as IEC/TR 62541 \cite{IECTR2010} by the \textit{International
Electrotechnical  Commission (IEC)}. Therefore, it enjoys widespread use.

\begin{table*}[ht!]
	\centering
	\caption{Threat Overview}
	\label{tab:threat}
	\resizebox{\textwidth}{!}{
		\begin{tabular}{|l|l|}
		\hline
		Potential Process Crash or Stop for Managed Device	&	Denial Of Service	\\
		\hline
		Data Flow Downstream Is Potentially Interrupted	&	Denial Of Service	\\
		\hline
		Potential Process Crash or Stop for smart energy controller	&	Denial Of Service	\\
		\hline
		Data Flow Upstream Is Potentially Interrupted	&	Denial Of Service	\\
		\hline
		Elevation Using Impersonation	&	Elevation Of Privilege	\\
		\hline
		Elevation Using Impersonation	&	Elevation Of Privilege	\\
		\hline
		Managed Device May be Subject to Elevation of Privilege Using Remote Code Execution	&	Elevation Of Privilege	\\
		\hline
		Elevation by Changing the Execution Flow in Managed Device	&	Elevation Of Privilege	\\
		\hline
		smart energy controller May be Subject to Elevation of Privilege Using Remote Code Execution	&	Elevation Of Privilege	\\
		\hline
		Elevation by Changing the Execution Flow in smart energy controller	&	Elevation Of Privilege	\\
		\hline
		Weak Authentication Scheme	&	Information Disclosure	\\
		\hline
		Downstream Data Flow Sniffing	&	Information Disclosure	\\
		\hline
		Upstream Data Flow Sniffing	&	Information Disclosure	\\
		\hline
		Potential Data Repudiation by Managed Device	&	Repudiation	\\
		\hline
		Potential Data Repudiation by smart energy controller	&	Repudiation	\\
		\hline
		Spoofing the smart energy controller Process	&	Spoofing	\\
		\hline
		Spoofing the Managed Device Process	&	Spoofing	\\
		\hline
		Spoofing the smart energy controller Process	&	Spoofing	\\
		\hline
		Replay Attacks	&	Tampering	\\
		\hline
		Collision Attacks	&	Tampering	\\
		\hline
		Potential Lack of Input Validation for Managed Device	&	Tampering	\\
		\hline
		Potential Lack of Input Validation for smart energy controller	&	Tampering	\\
		\hline
		\end{tabular}
	}
\end{table*}

\section{Security Analysis}
This section contains a threat model description to determine the security properties deemed necessary for the current
application, followed by an analysis of the OPC UA profiles for their general security and their features countering
the identified threats.

\subsection{Thread Model}
An adversary targeting the smart energy controller may have many potential goals (the list is non-exhaustive):
\begin{itemize}
	\item Extract information to draw conclusions on power consumptions, user behavior and billing information;
	\item Extract information to achieve credentials to administrative accounts;
	\item Manipulate values to achieve altered billings;
	\item Take over the device to alter power flows;
	\item Issue commands that stop the device from functioning;
	\item Manipulate values to evoke illegal conditions that damage the device.
\end{itemize}

Technique to achieve this can be categorized into the following (according to the \textit{STRIDE} methodology
\cite{potter2009microsoft}):
\begin{itemize}
	\item Spoofing of user identity;
	\item Tampering;
	\item Repudiation;
	\item Information disclosure (privacy breach or data leak);
	\item Denial of service (DoS);
	\item Elevation of privilege.
\end{itemize}

\begin{figure}[!t]
	\centering
		\scalebox{0.4}{
			\includegraphics{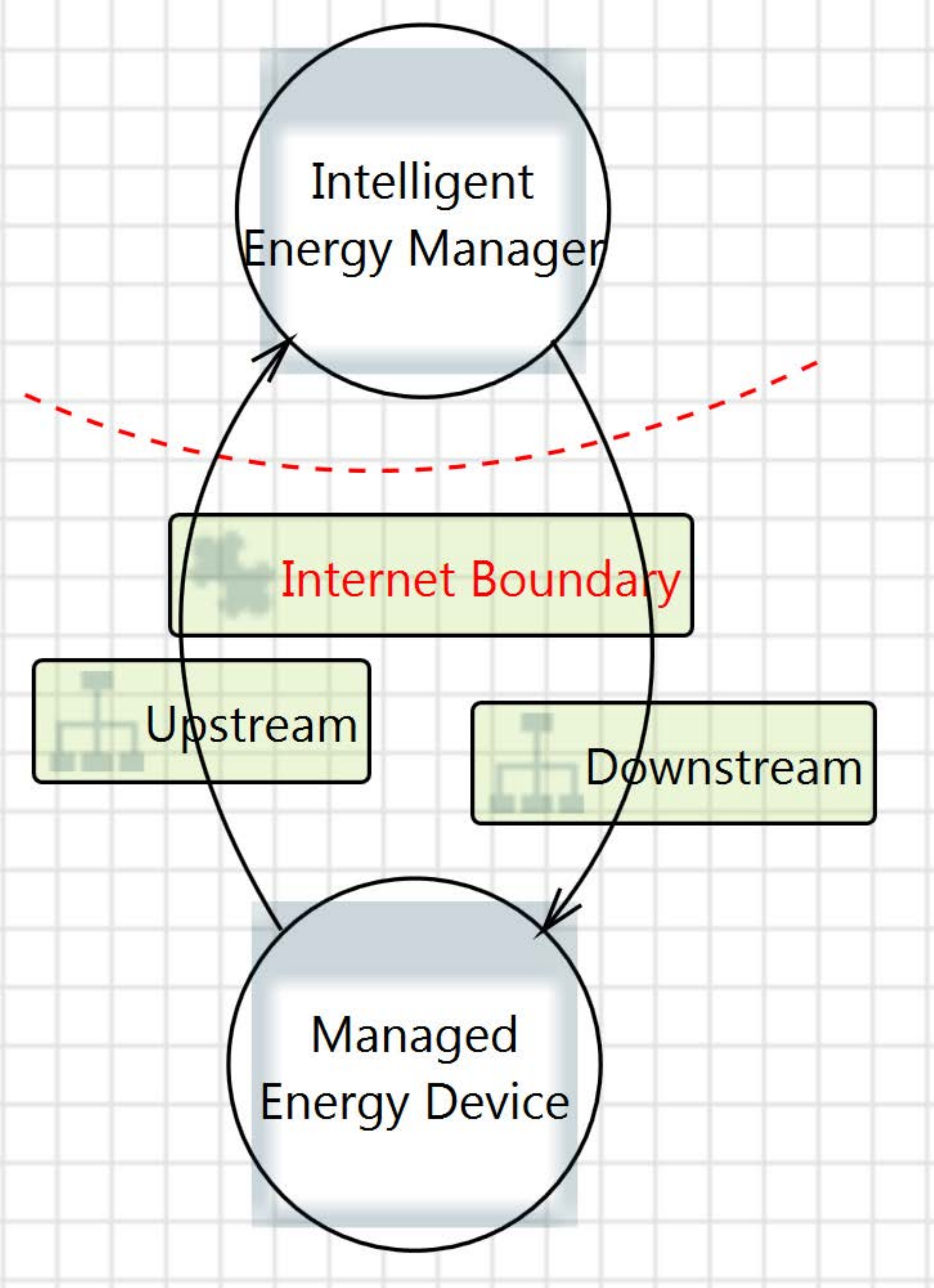}
		}
\caption{Graphic Threat Model Representation}
\label{fig:tm}
\end{figure}

To analyze OPC UA an unsecured bidirectional communication line between an assumed intelligent energy manager and a
managed device was modeled in the \textit{Microsoft Threat Modeling Tool 2016} \cite{potter2009microsoft}, using generic
data flows (see the graphical representation in Figure \ref{fig:tm}). This yielded in 5 DoS threats, 6 in privilege elevation, 3 in
information disclosure, 2 in repudiation, 3 in spoofing an 4 in tampering (see Table \ref{tab:threat} for details). The
next sections will reason on the possibility to counter these threats using OPC UA.  

\subsection{Analysis}
\label{sec:SA:OPC}

IoT connections for industrial usage should provide 
at minimum the same level of security as the IEC 62351 standard \cite{IEC:2007}.
This standard, however, does not necessarily assure end-to-end security \cite{Fries2010}.
Therefore, additional measures have to be taken to secure to provide a thoroughly secured connection.
As stated in Section \ref{sec:intro}, the OPC UA is deemed a suitable candidate to fulfill this requirement and moreover
provide end-to-end security. It provides three cipher suites, all of which use the \textit{Advanced
Encryption Standard (AES)} for symmetric encryption, \textit{Rivest-Shamir-Adleman (RSA)}-based asymmetric encryption with \textit{Optimal asymmetric
encryption padding (OAEP)}and \textit{Secure Hash Algorithm 2 (SHA2)}-based systems for message signing. All of these
suites have a signature length of 256 bits, channel nonce lengths of 32 bytes and asymmetric key lengths between 2048 and 4096 bits
, as well as they are using the \textit{Cipher Block Chaining (CBC)} mode of operation \cite{OPC2017}.
These combination of algorithms is currently
deemed secure \cite{NIST:2015}. The only exception to the algorithms stated above  is the use of the \textit{Secure Hash
Algorithm 1 (SHA1)} in the RSA-OAEP-SHA1 operation used for asymmetric encryption, which is, however, not regarded as a 
security issue, as RSA-OAEP only requires a hash function that has a neglectable possibility of producing all-zero
sequences, which SHA1 fulfills\footnote{These are, however, only the \textit{necessary}, not the \textit{sufficient}
security properties. There is yet no formal security proof for RSA-OAEP-SHA1.}\cite{cryptoeprint:2006:223}. The underlying RSA-OAEP possesses a formal security proof
\cite{Fujisaki2004}. A survey from the \textit{German Federal Office for Information Security (BSI)}
regarded \textsl{Basic256Sha256}  as the most secure of these suites \cite{Damm2017}. This, however, referred to a
previous version (1.2), whereas in the current (1.4) the \textsl{Aes256-Sha256-RsaPss} profile has been added. This
profile differs from the former in that it replaces SHA1 with SHA2 in asymmetric encryption and the \textit{RSA
Signature Algorithm Public-Key Cryptography Standards (RSASSA-PKCS)} version 1.5 with the
\textit{Probabilistic Signature Scheme (RSASSA-PSS)} for asymmetric signatures. As the latest version of the defining
standard \cite{RFC8017} both recommends using SHA2 and requires using RSASSA-PSS for robustness reasons, the newer
profile Aes256-Sha256-RsaPss is deemed superior over the one favored by the BSI.
Furthermore, the IEC standard from 2015 \cite{IECTR2015} defines TLS 1.0, 1.1 and 1.2 as means to secure communications,
although the newer technical specification by the OPC foundation only provides TLS 1.2 as single choice.
This standard provides three choices for TLS, AES256 with RSA key exchange or AES128 and AES256 with
\textit{Diffie-Hellman Ephemeral (DHE)} key exchange, each using SHA256 for hashing and the CBC mode.  Further
principally defined algorithms are AES-CTR (symmetric encryption), RSA-PKCS15 (asymmetric encryption) and
RSA-PKCS15-SHA1(signature) and P-SHA1 (key derivation), but these are not used in any profiles and therefore regarded
as if not supported \cite{OPC2017}.
Table \ref{tab:feat} provides an overview of the cryptographic algorithms used in the different OPC UA profiles.

In conclusion, using the Aes256-Sha256-RsaPss profile does mitigate the threats to information disclosure. It does also
mitigate collision attacks and should, by specification, prevent replay attacks in conjunction with a sequence number.
However, the BSI found out in its analysis that the provided reference implementation deviated from this specification,
as the sequence number was not evaluated \cite{Damm2017}. Other tampering-related attacks are in the devices' scopes rather
than in the protocol's. This is similar to privilege elevation attacks; impersonation attacks can be mitigated by
providing proper message authentication provided by the secure signatures in the profile, albeit, the other attacks
of this category are in the clients' scopes which also applies for client-based impersonation (which must be mitigated
by secure client authentication, e.g. via secure passwords and/or second factors). Also, spoofing of processes has to
countered on the client.
The threats regarding DoS  are infrastructural and, thus, need external counter
measures, except for message flooding, which can be partially mitigated by rate limiting through artificial delays \cite{Damm2017}.
The repudiation is not explicitly stated, but non-repudiation for sent messages is in general, as far the communications
protocol is concerned, provided by the usage of secure public key authentication and for received ones an issue of
thorough logging, which is out of the protocol's scope.
\begin{table*}[ht!]
	\centering
	\caption{Currently Valid OPC UA Security Profiles}
	\label{tab:feat}
	\resizebox{\textwidth}{!}{
		\begin{tabular}{|c|c|c|c|c|c|}
			\hline
			Name		&	Aes128-Sha256-RsaOaep	&	Basic256Sha256	&	Aes256-Sha256-RsaPss	&	TLS\_RSA\_AES\_256\_CBC\_SHA256	&	TLS\_DHE\_RSA\_AES\_nnn\_CBC\_SHA256\\
			\hline
			Symmetric Encryption	&	AES128-CBC	&	AES256-CBC	&	AES256-CBC	&	AES256-CBC	&	AES128/256-CBC	\\
			\hline
			Symmetric Signature	&	HMAC-SHA2-256	&	HMAC-SHA2-256	&	HMAC-SHA2-256	&	HMAC-SHA2-256	&	HMAC-SHA2-256	\\
			\hline
			Asymmetric Encryption	&	RSA-OAEP-SHA1	&	RSA-OAEP-SHA1	&	RSA-OAEP-SHA2-256	&	RSA	&	-	\\
			\hline
			Asymmetric Signature	&	RSA-PKCS15-SHA2-256	&	RSA-PKCS15-SHA2-256	&	RSA-PSS15-SHA2-256	&	RSASSA-PKCS15	&	RSASSA-PKCS15	\\
			\hline
			Certificate Signing	&	RSA-PKCS15-SHA2-256	&	RSA-PKCS15-SHA2-256	&	RSA-PKCS15-SHA2-256	&	RSASSA-PKCS15	&	RSASSA-PKCS15	\\
			\hline
			Key Derivation	&	P-SHA2-256	&	P-SHA2-256	&	P-SHA2-256	&	RSASSA-PKCS15	&	DHE	\\
			\hline

		\end{tabular}
	}
\end{table*}

\section{Conclusion and Outlook}
This paper showed that, in principle, OPC UA is suitable to serve as a secure architecture for a smart energy controller
and its managed devices, either standalone or by designing Modbus applications using OPC UA (for an example see
\cite{6566647}).
This allows for securing the application at a higher level, for with the most secure options being the
Aes256-Sha256-RsaPss profile or TLS with TLS\_DHE\_RSA with AES\_256\_CBC\_SHA256.
Complimentary to this, it is advisable to implement additional external measures to secure the solution against attacks
targeting the clients, provide rigorous logging facilities and implement external anti-flooding provisions.
This could be achieved by segregating the M2M networks from traditional ICT networks (and the Internet, except where a
\textit{Virtual Private Network (VPN)} runs over the former) and also from each other. This follows the principle of
least privilege, where an entity has only access to the particular resources it needs and adds an additional layer of
access protection as well as, through this, protection against flooding-based and other DoS attacks.
In general, the approach of defense-in-depth should be applied \cite{NSA:2010}, making the best effort to security on
any point of the system, for which using a secure distributed system architecture is the first step.

\bibliographystyle{IEEEtran} 
\bibliography{IEEEabrv,literature}

\end{document}